\documentclass[12pt]{article}
\def\lae{\;^{<}_{\sim} \;} \def\gae{\; ^{>}_{\sim} \;}

\title{\textbf{Triple Unification of Inflation, Dark matter and Dark energy in Chaotic Braneworld Inflation}}
{\author{\\[1cm]
{\sc \large Chia-Min Lin$^{\ast}$}\\
{\sl\small Department of Physics, National Tsing Hua University, Hsinchu, Taiwan 300 }
}}

\usepackage[margin=2cm]{geometry}
\usepackage{graphicx,color}
\usepackage{subfigure}
\begin{document}
\maketitle
\begin{abstract}
In this paper, we show that in the framework of chaotic braneworld inflation, after preheating, the remaining
oscillating inflaton field can play the role of dark matter with the observed level. Augmented by a non-zero effective cosmological constant $\Lambda_4$ on the brane, triple unification of inflation, dark matter and dark energy by a single field is realized. Our model perhaps is the simplest one in the market of theories to achieve triple unification.

\end{abstract}
\footnoterule{\small $^\ast$cmlin@phys.nthu.edu.tw}

\section{Introduction}
The almost scale invariant spectrum of Gaussian adiabatic density perturbations observed from Cosmic Microwave Background (CMB) by WMAP satellite support the idea of inflation as the standard paradigm for
the early universe. On the other hand, the "dark side of the universe" (dark energy and dark matter) currently dominate the energy budget of our universe. Usually we use different models to explain inflation, dark matter and dark energy, however in the interesting papers \cite{Liddle:2008bm, Liddle:2006qz}, it was suggested that inflation, dark matter and dark energy can be
unified by a single scalar field $\phi$ with the potential of the form
\begin{equation}
V=V_0+\frac{1}{2}m^2\phi^2,
\label{eq4}
\end{equation}
where $V_0$ is the dark energy we see today which may be chosen from the
string landscape, $\phi$ plays the role of an inflaton field during
inflation. This potential was also considered in \cite{Linde:2002ws} to unify inflation and dark energy.
After inflation, presumably there is a subsequent
preheating \cite{Kofman:1994rk} through a coupling $g\phi^2 \chi^2/2$ to some other scalar field
$\chi$, which makes the inflaton decays "incompletely". This reduces
the field value to $\phi=m/g$ and the remaining inflaton hopefully
will play the role of dark matter. Unfortunately, this
simple and elegant scenario does not work because $m$, which is fixed by CMB
temperature fluctuation, is too large so that unless $g=10^7$ \cite{Liddle:2006qz},
the reduced field value will be too large to make a successful dark matter.
Therefore extra mechanism to reduce the field value
is needed. For example, in \cite{Liddle:2008bm}, a subsequent
thermal inflation is assumed to further reduce the field value and in \cite{Cardenas:2007xh, Panotopoulos:2007ri}, the
authors suggest that plasma mass effects \cite{Kolb:2003ke} could provide the mechanism for an
incomplete reheating after inflation.

As shown in \cite{Maartens:1999hf}, the inflaton
mass $m$ in chaotic inflation can be suppressed to be much lower if
we consider the effect of modification to the Friedmann equation in
brane cosmology (an application of this result can be found in \cite{Bento:2004pz}).
Therefore it is worth to investigate whether $m$
can be small enough and the inflaton field value can be reduced to
the right value in order to play the role of dark matter with
reasonable value of $g$. In this paper, we show the answer is
positive. \emph{The point is conventional chaotic inflation plus preheating does NOT work for
a successful triple unification scenario, but chaotic braneworld inflation plus preheating simply works.}

The paper is organized as follows. In section \ref{2} We briefly review the basic
results of chaotic inflation on the brane which we call chaotic braneworld inflationin. In section \ref{sec2}, we present our main result and show why it can make a simple triple unification. Our conclusions are summarized in Section \ref{3}.
\section{Chaotic Braneworld Inflation}
\label{2}

In the braneworld inflation scenario where Einstein's equations hold in the five-dimensional bulk and
the matter fields are confined to the 3-brane, four-dimensional Einstein equations and Friedmann equation are modified \cite{Shiromizu:1999wj, Binetruy:1999hy, Flanagan:1999cu}. It is well known that the Friedmann equation in this
context is given by \cite{Maartens:1999hf}
\begin{equation}
H^2=\frac{8\pi}{3M_P^2}\rho\left(1+\frac{\rho}{2\lambda}\right),
\label{eq0}
\end{equation}
where $M_P=1.22 \times 10^{19}$ GeV is the 4D Planck mass and
$\lambda$ is the brane tension. Eq. (\ref{eq0}) reduces to the usual
Friedmann equation at $\rho \ll \lambda$. The Planck mass in four and five dimensions
is related by $\lambda$ via
\begin{equation}
M_P=\sqrt{\frac{3}{4\pi}}\frac{M_5^3}{\sqrt{\lambda}}.
\end{equation}
A lower bound of $M_5 \gae 10^8$ GeV $=8.2 \times 10^{-12}M_P$ is given by requiring the theory to reduce to Newtonian
gravity on scales larger than $1$ mm \cite{Maartens:1999hf}.

The curvature perturbation is given by
\begin{equation}
\zeta=\frac{H\delta\phi}{\dot{\phi}}.
\end{equation}
If the mass of the field $\phi$ is much smaller than Hubble parameter, the field fluctuations at horizon exit are given by $\langle\delta\phi^2\rangle \simeq (H/2\pi)^2$.
The spectrum of scalar perturbation at horizon exit is
\begin{equation}
A^2_s\equiv \frac{4}{25}\langle\zeta^2\rangle =\frac{512 \pi}{75 M^6_P}\frac{V^3}{V^{\prime 2}}\left(1+\frac{V}{2\lambda}\right)^3.
\label{eq3}
\end{equation}
In the slow-roll approximation, the total number of e-folds during inflation is given by
\begin{equation}
N \equiv \int^{t_f}_{t_i}Hdt \simeq -\frac{8\pi}{M^2_P}\int^{\phi_f}_{\phi_i}\frac{V}{V^{\prime}}\left(1+\frac{V}{2\lambda}\right) d\phi,
\end{equation}
where $i$ and $f$ are the values at the beginning and end of inflation,
respectively. The value $\phi_f$ is obtained from the condition $\max\{\epsilon, |\eta|\}=1$, where $\epsilon$ and $\eta$ are the slow-roll parameters, given by
\begin{equation}
\epsilon \equiv \frac{M^2_P}{16\pi}\left(\frac{V^\prime}{V}\right)^2\frac{1+V/\lambda}{(1+V/2\lambda)^2},
\end{equation}
\begin{equation}
\eta \equiv \frac{M^2_P}{8\pi}\frac{V^{\prime\prime}}{V}\frac{1}{1+V/2\lambda}.
\end{equation}
Using Eq. (\ref{eq4}) and (\ref{eq3}) and imposing CMB normalization $A_s(\phi(N=60)) \approx 2 \times 10^{-5}$ \cite{Komatsu:2008hk}, we obtain
\begin{equation}
m \approx 4.5 \times 10^{-5} M_5.
\label{eq1}
\end{equation}
This result is valid for $V/\lambda \gg 1$ which gives an upper bound on $M_5$, namely, $M_5 \ll 10^{17}$ GeV $=8.2 \times 10^{-3}M_P$. The $V_0$ in Eq. (\ref{eq0}) is the effective cosmological constant $\Lambda_4$ on the brane which is given by
\begin{equation}
V_0=\Lambda_4=\frac{4\pi}{M^3_5}\left(\Lambda+\frac{4\pi}{3M_5}\lambda^2\right).
\end{equation}

\section{Triple Unification}
\label{sec2}

After inflation, the scalar field start to oscillate at $t_\ast$,
for $t>t_\ast$, we have \footnote{In brane inflation scenario, the continuity equation is not changed \cite{Binetruy:1999ut}. }
\begin{equation}
\rho_\phi=\frac{1}{2}m^2\phi^2_\ast \left(\frac{a_\ast}{a}\right)^3,
\end{equation}
and the radiation evolves as
\begin{equation}
\rho_R=\rho^\ast_R \left(\frac{a_\ast}{a}\right)^4,
\end{equation}
where $\rho^\ast_R=3M_P^2 m^2/8 \pi$. We assume here an adiabatic
expansion where the entropy $S=gT^3a^3$ is constant during the evolution and $g$
is the entropic degrees of freedom. Then the number density of
radiation is
\begin{equation}
n_{\gamma,0}=n^\ast_\gamma \frac{g_\ast}{g_0} \left(\frac{a_\ast}{a_0}\right)^3,
\end{equation}
where the zero subscript indicates current values. By using
$\rho^\ast_R=\pi^2 g_\ast T^4/30$ and $n_\gamma=2\zeta(3)T^3/\pi^2$,
we can obtain the current dark matter per photon ratio as \cite{Liddle:2008bm}
\begin{equation}
\xi_{dm,0} \equiv \frac{\rho_{\phi,0}}{n_{\gamma,0}} \simeq 4 \frac{g_0}{g^{1/4}_\ast}
\left(\frac{m}{M_P}\right)^{1/2}\frac{\phi^2_\ast}{M_P^2}M_P.
\end{equation}
We can calculate $n_{\gamma,0}$ from current Cosmic Microwave
Background (CMB) black body temperature and $\rho_{\phi,0}$ from the cold dark
matter energy density $\Omega_c$, hence a value
$\xi_{dm,0}=2.2\times 10^{-28}M_P$ can be obtained from observation
, which for typical values $g_\ast \sim 100$ and $g_0=3.9$ gives \cite{Komatsu:2008hk}
\begin{equation}
\left(\frac{m}{M_P}\right)^{1/2}\frac{\phi^2_\ast}{M_P^2} \simeq 4 \times 10^{-29}.
\label{eq2}
\end{equation}

\begin{figure}[htbp]
\begin{center}
\includegraphics[width=0.5\textwidth]{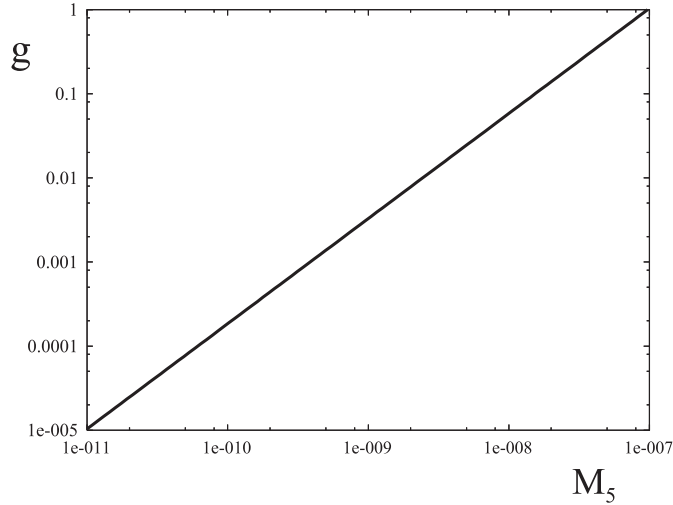}
\caption{$g$ versus $M_5$ with unit $M_P$}
\label{fig1}
\end{center}
\end{figure}

After preheating, $\phi=m/g$. We demand
\begin{equation}
\phi=\phi_\ast.
\end{equation}
If the above condition is satisfied, naturally after preheating the original inflaton field will play the role of dark matter model as Eq. (\ref{eq2}) is satisfied, \emph{this is also necessary in order to permit a long radiation-dominated epoch and where the "conventional chaotic inflation plus preheating" scenario fails}, because a too large $\phi$ is obtained after preheating and a further reduction of the field amplitude is required. However, in our case by using Eq. (\ref{eq1}) and (\ref{eq2}), we obtain
\begin{equation}
g=5.83 \times 10^8 (M_5/M_P)^{5/4}.
\end{equation}
The result is clearly shown in Fig. (\ref{fig1}). As we can see, there is a regime $10^{-11} \lae M_5/M_P \lae 10^{-7}$ where
we have $10^{-5} \lae g \lae 1$. This is a perfectly reasonable range both for conventional preheating or brane preheating \cite{Tsujikawa:2000hi}\footnote{Our result do not depend on which kind of preheating we use.}. Therefore in our case, no further reduction of $\phi$ after preheating is needed. After preheating, our universe enters the era of "hot big bang" via reheating and becomes radiation dominated and the remaining oscillating inflaton field becomes dark matter until now.

\section{Conclusion}
\label{3}
We have shown that triple unification of inflation, dark matter and dark energy can simply be achieved in the
framework of chaotic braneworld inflation plus preheating. The required 5D Planck mass is in the range $10^{-11} \lae M_5/M_P \lae 10^{-7}$ corresponds to the coupling for reheating $10^{-5} \lae g \lae 1$. This model is very simple (perhaps the simplest one among all theories of triple unification) but it elegantly realized the idea of triple unification because no further reduction mechanism (for example, thermal inflation as in \cite{Liddle:2008bm}) for the inflaton field $\phi$ is needed.



\section*{Acknowledgement}
This work is supported by the NSC under grant No. NSC 96-2628-M-007-002-MY3, by the NCTS, by the Boost Program of NTHU and by the KEK exchange program. The author would like to thank KEK for hospitality during this work.


\end{document}